\documentclass{article}

\usepackage{arxiv}

\usepackage[utf8]{inputenc} 
\usepackage[T1]{fontenc}    
\usepackage[hidelinks]{hyperref}       
\usepackage{url}            
\usepackage{booktabs}       
\usepackage{amsfonts}       
\usepackage{nicefrac}       
\usepackage{microtype}      
\usepackage{lipsum}
\usepackage{graphicx}
\graphicspath{ {./images/} }
\usepackage{tikz}
\usetikzlibrary{arrows, positioning, calc}

\title{Mind the Gap: How the Technical Mechanisms of Agentic AI Outpace Global Legal Frameworks}

\author{
 Marcel Osmond \\
  Centre for Commercial Law Studies (CCLS)\\
  Queen Mary University of London\\
    Postgraduate Researcher\\
  London, WC2A 3JB, United Kingdom \\
  \texttt{m.osmond@hss25.qmul.ac.uk} \\
  \and
 \textbf{Thomas Jego} \\
  DeVinci Higher Education \\
  L\'eonard de Vinci Graduate School of Engineering\\
    Postgraduate Researcher\\
  Paris - La D\'efense, 92400, France \\
  \texttt{thomas.jego@edu.devinci.fr} \\
}

\begin{document}

\maketitle
\begin{abstract}
This article presents the first systematic comparative survey of how public bodies, international organisations, national regulators, and the private sector define agentic artificial intelligence, identifying the technical inaccuracies pervading each definition. Analysing eleven regulatory instruments and industry frameworks --- including the EU AI Act, the OECD/G7 Principles, NIST, the UK ICO, and the European Commission --- alongside six leading developer architectures, this study demonstrates a persistent definitional gap: legal definitions consistently conflate model capability with agentic architecture, attribute cognitive deliberation to probabilistic token prediction, and treat autonomy as a scalar property rather than a structural shift from single-inference to iterative execution loops with tool integration. A consensus technical definition synthesised from developer documentation is proposed. The article examines the consequences of this gap, demonstrating that definitional imprecision produces regulatory instruments structurally incapable of governing the actual mechanisms --- system prompts, API permissions, sandboxing, and orchestration code --- that constitute agentic autonomy.
\end{abstract}

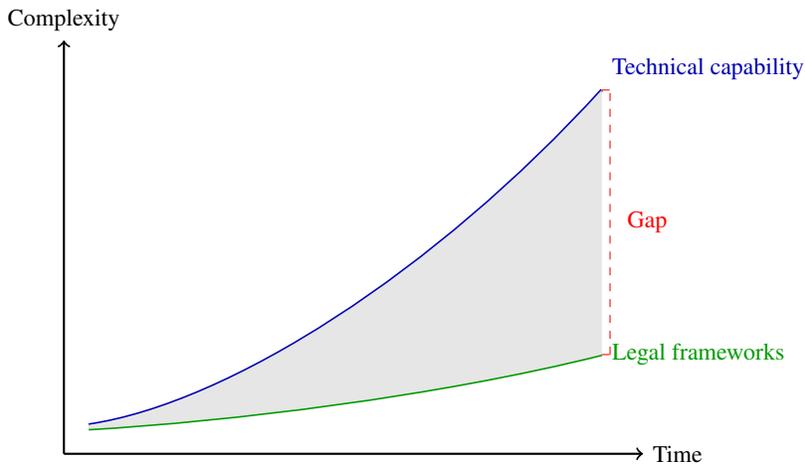
\begin{figure}[!ht]
\centering
\begin{tikzpicture}[
    scale=1.1,
    axis/.style={thick},
    line/.style={very thick},
    every node/.style={font=\small},
    gapline/.style={red, dashed, thin},
]

\draw[axis, ->] (0,0) -- (7,0) node[right] {Time};
\draw[axis, ->] (0,0) -- (0,5) node[above] {Complexity};

\draw[line, green!60!black] 
(0.3,0.3)
.. controls (2,0.4) and (4.5,0.7)
.. (6.5,1.2)
node[right] {Legal frameworks};

\draw[line, blue!70!black]
(0.3,0.35)
.. controls (2,0.6) and (4.5,2.2)
.. (6.5,4.4)
node[above right] {Technical capability};

\draw[gapline] (6.5,1.2) -- (6.6,1.2);
\draw[gapline] (6.5,4.4) -- (6.6,4.4);
\draw[gapline] (6.6,1.2) -- (6.6,4.4);

\fill[gray!20]
(0.3,0.3)
.. controls (2,0.4) and (4.5,0.7)
.. (6.5,1.2)
--
(6.5,4.4)
.. controls (4.5,2.2) and (2,0.6)
.. (0.3,0.35)
-- cycle;

\node[red] at (7.05,2.8) {Gap};

\end{tikzpicture}
\caption{Conceptual visualization of the widening gap between technological capability growth and regulatory development.}
\label{fig:gap_area}
\end{figure}

\keywords{Agentic AI \and AI Regulation \and Definitional Gap \and EU AI Act \and Autonomous Systems \and AI Governance}

\newpage
\section{Introduction}

Artificial intelligence systems capable of autonomous, multi-step action --- systems that can browse the web, execute code, query databases, send emails and interact with external services without human confirmation at each step --- are no longer a theoretical prospect. They are being deployed commercially, integrated into professional workflows, and embedded in consumer products at a pace that has consistently outrun the regulatory frameworks nominally designed to govern them. This paper is concerned with a specific and foundational dimension of that governance gap: the failure of legal and policy definitions to accurately capture what agentic AI systems are, how they work, and where the real locus of risk and responsibility lies.

The definitional question is not a preliminary technicality to be resolved before the serious legal analysis begins. It is the serious legal analysis. A regulatory framework that mischaracterises its subject matter will, by construction, generate compliance obligations calibrated to the wrong features of the technology. It will direct liability toward entities that do not control the relevant risks and away from those that do. It will require testing and certification procedures that cannot meaningfully be applied to systems whose behaviour is non-deterministic and dynamically generated. And it will create safe harbours --- whether intentionally or not --- for the specific architectural decisions that actually determine whether an agentic system causes harm.

This paper demonstrates, through a systematic comparative survey of eleven regulatory instruments and policy frameworks, that this failure is not isolated or incidental. It is structural, recurrent, and traceable to identifiable methodological causes. The frameworks examined span four continents and include binding legislation, intergovernmental guidance, national regulatory authority publications and private sector documentation: the EU Artificial Intelligence Act; the European Data Protection Supervisor's TechSonar guidance; the European Commission's Agentic AI Landscape Report; the OECD Recommendation on Artificial Intelligence and the G7 Guiding Principles built upon it; the NIST AI Risk Management Framework; the UK Information Commissioner's Office guidance on agentic AI; the Singapore Model AI Governance Framework; the New Zealand Ministry of Business, Innovation and Employment's Responsible AI Guidance; the New South Wales Government's AI Agent Usage and Deployment Guidance; and the published academic framework of Maas and Olasunkanmi on Treaty-Following AI. These instruments were selected on the basis of their direct definitional engagement with autonomous or agentic AI systems and their significance within their respective regulatory traditions.

To evaluate the technical accuracy of each definition, the paper draws on the developer documentation of six leading agentic AI developers --- Anthropic, Google Cloud, OpenAI, Microsoft, C3.ai, and Meta --- extracting from those sources the elements of agentic architecture on which technical consensus is clearest. This methodology grounds the critical analysis in the operational reality of deployed systems rather than in theoretical computer science abstraction, and produces, as a secondary contribution, a synthesised technical definition that regulatory drafters could adopt or adapt.

The paper proceeds in three parts. Part~I surveys each of the eleven frameworks in turn, identifying in each case the specific technical inaccuracy or architectural misunderstanding that produces what this paper terms the \textit{definitional gap}. Part~II synthesises these findings into four recurring categories of definitional failure --- anthropomorphism, architectural conflation, the levels-of-autonomy fallacy, and the learning-and-memory myth --- and examines the structural reasons why non-technical authorship processes systematically produce them. The conclusion draws out the legal consequences of the definitional gap and identifies the conditions for more technically adequate regulatory instruments.

The paper's original contribution is threefold. It is the first systematic comparative survey of agentic AI definitions across this range of jurisdictions and institutional contexts. It identifies and names four specific categories of definitional failure that recur across the corpus regardless of legal tradition or institutional origin. And it demonstrates that the consequence of these failures is a body of regulatory instruments that are simultaneously overbroad and underinclusive --- reaching AI systems that pose no distinctive agentic risk while failing to govern the specific engineering decisions that determine the real-world risk profile of agentic deployments.

\newpage
\section{Legal and Policy Definitions of Agentic AI: A Comparative Survey and Technical Gap Analysis}
\label{sec:survey}

There is currently no widely adopted, binding legal definition of Agentic AI in major treaties, statutes, or regulations. Existing laws generally regulate these systems under broader ``AI system,''\footnote{Andy Goldstein, Agentic AI, European Data Protection Supervisor. Accessible at: \url{https://www.edps.europa.eu/data-protection/technology-monitoring/techsonar/agentic-ai_en}} automated system, or sectoral concepts rather than using ``agentic'' as a defined legal term. The following survey examines the principal definitions advanced by public bodies, international frameworks, national regulators, and the private sector, and identifies, in each case, the technical inaccuracies and architectural misunderstandings that produce what this paper terms the \textit{definitional gap}.

Synthesising the technical documentation of six leading agentic AI developers --- Anthropic, Google Cloud, OpenAI, Microsoft, C3.ai, and Meta --- and extracting those elements on which technical consensus is clearest, the following definition is proposed:

\begin{quote}
``An agentic AI system is an autonomous computational system that pursues defined goals by autonomously perceiving context, planning, and executing multi-step actions --- through access to tools, external systems, and other agents --- with minimal human intervention once objectives are set, while adapting its behaviour based on human feedback and operating within prescribed instructions and guardrails.''\footnote{Google Cloud, Agentic AI (Google Cloud Documentation) available at: \url{https://cloud.google.com/discover/what-is-agentic-ai}}\textsuperscript{,}\footnote{Anthropic, Empirical Work on Agents (Anthropic Research) available at: \url{https://www.anthropic.com/research/measuring-agent-autonomy}}\textsuperscript{,}\footnote{OpenAI, Developer Documentation on Agents (OpenAI Developer Platform) available at: \url{https://developers.openai.com/tracks/building-agents/}}\textsuperscript{,}\footnote{Microsoft, Agent Factory: The new era of agentic AI (Microsoft Azure) available at: \url{https://azure.microsoft.com/en-us/blog/agent-factory-the-new-era-of-agentic-ai-common-use-cases-and-design-patterns/}}\textsuperscript{,}\footnote{C3.ai, Agentic System Architecture (C3.ai Documentation) available at: \url{https://c3.ai/blog/agentic-ai-explained/}}\textsuperscript{,}\footnote{Meta, AI Security Research (Meta AI Research) available at: \url{https://ai.meta.com/blog/practical-ai-agent-security/}}
\end{quote}
By defining agentic AI as Level~4 or~5 systems, regulators effectively create a \textit{de facto} safe harbour for Level~3 autonomous agents --- systems that use a single model to plan and execute multi-step actions without persistence, and which already possess the tool integration and iterative execution loop that constitute genuine agentic risk. This misclassification is a direct consequence of the definitional failures examined in the survey that follows.

\subsection{The European Data Protection Supervisor (EDPS)\protect\footnote{Andy Goldstein (n~1).}}
\label{subsec:edps}

\subsubsection{Definition}

So far, the clearest quasi-official description comes from the European Data Protection Supervisor, though it bears emphasis that this constitutes a technical and policy description rather than a binding legal definition in an act or treaty.\footnote{Andy Goldstein (n~1).} The EDPS defines Agentic AI as an autonomous system designed to fulfil complex, overarching goals rather than isolated tasks, without step-by-step human instruction. It is characterised by its ability to reason and plan, utilise external tools --- including APIs, databases, and web search --- and retain persistent memory to learn, adapt, and self-correct errors. Crucially, the EDPS framework distinguishes ``Agentic AI'' from simple ``AI Agents,'' defining Agentic AI specifically as an orchestrator that coordinates multiple, communicating sub-agents to achieve larger objectives.\footnote{Andy Goldstein (n~1).}

\subsubsection{Technical Gap}

The EDPS TechSonar framework reserves the label ``Agentic AI'' exclusively for an orchestrator coordinating multiple sub-agents, treating a single autonomous agent as something categorically different and lesser. This is technically inaccurate. As Anthropic's empirical work establishes, tool use is ``the building blocks of agent behaviour,''\footnote{Anthropic (n~3)} meaning that a single AI system equipped with tools and pursuing goals autonomously already qualifies as an agentic AI system --- no multi-agent coordination is required.

Google Cloud similarly defines agentic AI primarily around ``autonomous decision-making and action'' by a system that can ``set goals, plan, and execute tasks,''\footnote{Google Cloud (n~2)} without requiring orchestration of sub-agents as a definitional threshold.

Multi-agent coordination is therefore best understood as an advanced architectural feature of some agentic systems, as acknowledged by Google and C3.ai,\footnote{C3.ai (n~6)} rather than the defining criterion of agentic AI itself.

\subsection{The EU Artificial Intelligence Act}
\label{subsec:euaiact}

\subsubsection{Definition}

Because the EU AI Act does not explicitly define Agentic AI, it regulates agentic systems by grouping them into broad, catch-all definitions. Under Article~3,\footnote{Regulation (EU) 2024/1689 of the European Parliament and of the Council of 13 June 2024 laying down harmonised rules on artificial intelligence [2024] OJ L 2024/1689, Art 3(1). Available at: \url{https://eur-lex.europa.eu/legal-content/EN/TXT/HTML/?uri=OJ:L_202401689}} the Act treats an agentic system as an AI System: a machine-based system designed to operate with ``varying levels of autonomy'' and ``adaptiveness,'' which infers how to generate outputs --- including predictions and decisions --- that influence physical or virtual environments based on ``explicit or implicit objectives.''\footnote{ibid Art 3(1).} The Act further regulates these systems based on their ``Intended Purpose'' and assigns liability to two rigid roles: the ``Provider,'' who develops the system, and the ``Deployer,'' who uses it.\footnote{ibid Art 3(4).}

\subsubsection{Technical Gap}

The EU AI Act's foundational flaw is its reliance on a static ``intended purpose'' and its concept of ``reasonably foreseeable misuse.'' From a technical perspective, Agentic AI relies on an iterative execution loop and tool integration to dynamically generate novel, unprogrammed paths to achieve an objective. By definition, the specific steps an agent takes are non-deterministic, making all intermediate actions inherently unforeseeable by the original developer. Furthermore, the Act's characterisation of AI systems as having ``varying levels of autonomy''\footnote{ibid Art 3(1).} demonstrates a profound mechanical misunderstanding: in computer science, agentic autonomy is not a ``level'' or a slider. It is the architectural shift from a single inference pass to a continuous ReAct (Reason, Act, Observe) loop --- a structural difference, not a scalar one. Finally, the rigid legal separation between a ``provider'' and a ``deployer'' collapses in composite agentic architectures. If a deployer takes an open-weight LLM built by a provider, wraps it in a LangChain execution loop, and grants it API access to a live database, that deployer has fundamentally altered the system's agency. The Act's definitions provide no clear framework for attributing liability for the emergent behaviours of this composite architecture.

\subsection{The G7 and OECD}
\label{subsec:g7oecd}

\subsubsection{Definition}

The G7 Guiding Principles, building directly upon the OECD's framework, do not recognise ``Agentic AI'' as a distinct legal or technical category. Instead, they fold agent-like autonomy into two broader definitions. First, the OECD defines an AI System as ``a machine-based system that, for explicit or implicit objectives, infers, from the input it receives, how to generate outputs such as predictions, content, recommendations, or decisions that can influence physical or virtual environments,''\footnote{OECD, `Recommendation of the Council on Artificial Intelligence' (OECD Legal Instruments, OECD/LEGAL/0449, 2024). Available at: \url{https://legalinstruments.oecd.org/en/instruments/OECD-LEGAL-0449}} and the framework assumes these systems ``vary in their levels of autonomy and adaptiveness after deployment.''\footnote{ibid.} Second, the G7 specifically targets what it calls ``Advanced AI,'' defining it simply as ``the most advanced foundation models and generative AI systems.''\footnote{White \& Case, `AI Watch: Global regulatory tracker --- G7' (White \& Case, 4 December 2024). Available at: \url{https://www.whitecase.com/insight-our-thinking/ai-watch-global-regulatory-tracker-g7}}

\subsubsection{Technical Gap}

The foundational friction within the G7 and OECD frameworks is their conflation of model capability with agentic architecture. By defining the systems requiring the most scrutiny as ``the most advanced foundation models and generative AI systems,''\footnote{Ibid.} the G7 fundamentally mischaracterises how agency works. In computer science, agency is not determined by how ``advanced'' a foundation model is. A massive, state-of-the-art LLM used purely for zero-shot text generation has zero agency, whereas a smaller, older model integrated into an iterative execution loop with database write-access is highly agentic. Furthermore, the G7 Code of Conduct's warning against models that are ``self-replicating''\footnote{ibid.} heavily anthropomorphises the technology, treating neural network weights as biological viruses. A model cannot mechanically self-replicate; it can only generate code scripts that, if executed by a poorly sandboxed external environment, might copy files. The G7 seeks to govern the generative AI systems themselves, rather than regulating the programmatic wrappers, tool-calling permissions, and continuous loops that actually construct autonomous agency.

\subsection{The National Institute of Standards and Technology (NIST)}
\label{subsec:nist}

\subsubsection{Definition}

Because the NIST AI Risk Management Framework does not separately define ``agentic AI,'' it regulates agentic architectures under its catch-all definition of standard AI systems. NIST defines an AI system as ``an engineered or machine-based system that can, for a given set of objectives, generate outputs such as predictions, recommendations, or decisions influencing real or virtual environments.''\footnote{National Institute of Standards and Technology (NIST), `Artificial Intelligence Risk Management Framework (AI RMF 1.0)' (NIST AI 100-1, US Department of Commerce, January 2023) 1. Available at: \url{https://nvlpubs.nist.gov/nistpubs/ai/NIST.AI.100-1.pdf}} The framework states that these systems ``are designed to operate with varying levels of autonomy,''\footnote{ibid 6.} noting that ``human-AI configurations can span from fully autonomous to fully manual.''\footnote{ibid 45.} To manage the risks of these systems, NIST relies heavily on ``test, evaluation, verification, and validation (TEVV)'' processes throughout the system's lifecycle.\footnote{ibid 14.}

\subsubsection{Technical Gap}

The primary friction within the NIST AI Risk Management Framework is its conceptualisation of autonomy as a sliding scale and its focus on static outputs rather than continuous execution. From a computer science perspective, Agentic AI is not merely a standard predictive model with the human ``level'' dialed down; it is a distinct, composite architecture built on an iterative execution loop and tool integration. An agent does not simply output a ``recommendation''; it translates its reasoning into executable code to trigger external APIs without human review. Consequently, NIST's heavy reliance on ``test, evaluation, verification, and validation (TEVV)''\footnote{ibid 14.} breaks down when applied to Agentic AI. Traditional TEVV is designed to measure the accuracy of a static output against a known baseline. It cannot reliably validate a system that dynamically generates non-deterministic, multi-step execution paths to achieve an objective. By treating agency as a mere feature of automation rather than a fundamental architectural shift, NIST provides a risk management framework that is structurally inadequate for auditing true agentic mechanisms.

\subsection{The UK Information Commissioner's Office (ICO)}
\label{subsec:ico}

\subsubsection{Definition}

The UK ICO explicitly acknowledges that there is no formal legal definition for Agentic AI.\footnote{Slaughter and May, `ICO publishes report on agentic AI and its data privacy implications' (The Lens, 2026). Available at: \url{https://thelens.slaughterandmay.com/post/102me9n/ico-publishes-report-on-agentic-ai-and-its-data-privacy-implications}} In its early-stage guidance, the ICO describes Agentic AI as a system that ``integrates large language models with other tools''\footnote{Ibid.} and is capable of working with various inputs to ``plan, reason, take actions and learn.''\footnote{ibid.} The ICO highlights the technology's capacity for ``increased levels of autonomy'' and its ability to ``make decisions and take actions independently,''\footnote{Information Commissioner's Office, `AI'll get that! Agentic commerce could signal the dawn of personal shopping `AI-gents'' (ICO News and Blogs, 8 January 2026). Available at: \url{https://ico.org.uk/about-the-ico/media-centre/news-and-blogs/2026/01/ai-ll-get-that/}} such as acting as a ``personal shopping `AI-gent'.''\footnote{Ibid.} These systems are characterised by their ability to automate activities by ``interacting with its environment, solving problems in real time and mimicking some types of reasoning and planning.''\footnote{ibid.} The ICO warns of novel risks, including the ``purposes of an agentic AI system being too wide,'' ``a system being given access to personal data beyond what is necessary,'' and the potential for ``rapid automation of tasks which could result in an increase in automated decision-making (ADM).''\footnote{Slaughter and May (n~28).}

\subsubsection{Technical Gap}

While the ICO accurately identifies the integration of LLMs with external tools, its framework suffers from anthropomorphic friction. An LLM does not perform cognitive reasoning; it executes probabilistic inference. This misunderstanding leads directly to the regulatory fear of a system's purposes ``being too wide.''\footnote{ibid.} In a deterministic computing environment, an agentic system's purpose cannot be ``too wide'' independently of its programming; it is strictly bound by the system prompt and the specific API tools it is granted access to. If an agent accesses personal data ``beyond what is necessary,''\footnote{ibid.} it is not because the AI spontaneously broadened its own purpose, but because the human developer failed to implement granular access controls and proper sandboxing for the execution loop. The ICO's focus on the AI ``mimicking some types of reasoning''\footnote{Information Commissioner's Office (n~31).} obscures the actual regulatory target: the developer's architectural decisions regarding tool permissions and data access.

\subsection{Maas and Olasunkanmi}
\label{subsec:maas}

\subsubsection{Definition}

In their working paper, Maas and Olasunkanmi define AI agents as ``systems which can be instructed in natural language and then act autonomously,'' noting they are ``capable of pursuing difficult goals in complex environments without detailed (follow-up) instruction,'' through the use of ``affordances or design patterns such as tool use (e.g., web search) or planning.''\footnote{Matthijs M Maas and Tobi Olasunkanmi, `Treaty-Following AI' (2025) Institute for Law \& AI Working Paper Series No. 1-2025, 5. Available at: \url{https://law-ai.org/wp-content/uploads/2025/12/Maas-Olasunkanmi-2025-Treaty-Following-AI-final-LawAI-WPS-1-2025-4.pdf}} Building on this, they introduce the concept of Treaty-Following AI (TFAI), defined as ``agentic AI systems that are designed to generally follow their principals' instructions loyally but to refuse to take actions that violate the terms and obligations of a designated referent treaty text.''\footnote{ibid 14.} To achieve this, the authors propose a ``treaty-interpreting chain-of-thought loop'' where the AI dedicates computing time to ``reflect on potential treaty issues'' and ``reach a decision over the legality of its goals or actions'' before executing them.\footnote{ibid 32.}

\subsubsection{Technical Gap}

While Maas and Olasunkanmi accurately recognise the role of tool integration and iterative execution loops, their framework suffers from profound cognitive friction regarding the probabilistic nature of the foundation model. The authors propose a ``treaty-interpreting chain-of-thought loop,'' assuming the AI can ``reflect'' on legal issues and ``reach a decision'' to refuse an illegal action.\footnote{ibid 32.} An LLM does not cognitively weigh legal risk like a human judge; it performs a statistical token-prediction pass based on its system prompt. Relying on an LLM's chain-of-thought reasoning to self-police its own API calls is not a structural constraint; it is merely advanced prompt engineering. The authors attempt to solve a deterministic architecture problem --- controlling what external functions an agent can trigger --- with a probabilistic text-generation workaround, hoping the agent outputs a refusal token rather than an execute token. By treating the model's internal inference steps as actual legal deliberation, the TFAI framework conflates the generation of legally-sounding text with actual, hardcoded access controls through Identity and Access Management and sandboxing.

\subsection{The European Commission Agentic AI Landscape Report}
\label{subsec:ecreport}

\subsubsection{Definition}

The European Commission's report defines Agentic AI as ``an advanced form of artificial intelligence focused on autonomous decision-making and action'' that can ``set goals, plan, and execute tasks with minimal human intervention.''\footnote{European Commission, `European agentic AI landscape' (2025) 4. Available at: \url{https://digital-strategy.ec.europa.eu/en/library/agentic-ai-leveraging-european-ai-talent-and-regulatory-assets-scale-adoption}} The report asserts a specific architectural definition, stating that Agentic AI is ``a multi-agent system in which each agent performs a specific subtask required to reach a goal, with their efforts coordinated through AI orchestration.''\footnote{ibid 4.} It attributes cognitive-sounding capabilities to these systems, including ``Reasoning,'' which it defines as ``Contextual decision-making to make judgment calls and weigh tradeoffs.''\footnote{ibid.}

\subsubsection{Technical Gap}

The European Commission's landscape report suffers from two profound definitional frictions: architectural conflation and cognitive anthropomorphism. The report's assertion that Agentic AI \textit{is} a ``multi-agent system''\footnote{ibid 4.} is a categorical error. ``Agentic'' is a behavioural property achieved via an iterative execution loop and tool integration; a single LLM operating autonomously in a ReAct loop is highly agentic. By defining the technology by the number of models communicating rather than the mechanisms of their execution, the framework fundamentally misrepresents how agency is achieved. Furthermore, the report's claim that these systems use ``Reasoning'' to make ``judgment calls and weigh tradeoffs''\footnote{ibid.} elevates probabilistic token prediction to conscious deliberation. An LLM does not make a ``judgment call''; it performs multi-pass probabilistic token prediction to satisfy a predefined reward function or system prompt. By using the language of human deliberation, the report invites policymakers to regulate the AI's non-existent judgment rather than focusing on the deterministic access controls, data boundaries, and API permissions that actually govern the system's external actions.

\subsection{The Singapore Model AI Governance Framework}
\label{subsec:singapore}

\subsubsection{Definition}

Because the framework does not explicitly define ``Agentic AI,'' it regulates autonomous agentic architectures through a combination of its general AI definition and its classification of ``Human-out-of-the-loop'' operational models. The framework defines AI as technologies ``that seek to simulate human traits such as knowledge, reasoning, problem solving, perception, learning and planning,''\footnote{Personal Data Protection Commission Singapore (PDPC), `Model Artificial Intelligence Governance Framework (Second Edition)' (2020) 18. Available at: \url{https://www.pdpc.gov.sg/-/media/files/pdpc/pdf-files/resource-for-organisation/ai/sgmodelaigovframework2.pdf}} which ``produce an output or decision (such as a prediction, recommendation, and/or classification).''\footnote{ibid 18.} To manage systems that act independently, the framework categorises them under a ``Human-out-of-the-loop''\footnote{ibid 30.} approach, defined as situations where ``there is no human oversight over the execution of decisions,''\footnote{ibid.} and where the ``AI system has full control without the option of human override.''\footnote{ibid.}

\subsubsection{Technical Gap}

The primary definitional friction within the Singapore Model Framework stems from its anthropomorphic baseline and its conceptualisation of operational control. An LLM simulates none of the cognitive traits --- knowledge, reasoning, problem solving, perception, learning, planning --- that the framework ascribes to it; it performs probabilistic token prediction. This cognitive framing leads to a flawed regulatory mechanism. The framework treats autonomy merely as a ``Human-out-of-the-loop'' scenario, concluding that the ``AI system has full control''\footnote{ibid.} of its executions. An AI system never has ``full control'' in the human, deliberative sense. Autonomy is not achieved by granting the machine free will, but by architecting an iterative execution loop that programmatically triggers external tools and APIs. By stating the AI has ``full control,'' the framework shifts the regulatory focus toward the model's independent behaviour, rather than targeting the actual mechanism of agency: the deterministic boundaries, system prompts, and API permissions explicitly coded by the developer.

\subsection{The New Zealand Ministry of Business, Innovation and Employment (MBIE)}
\label{subsec:nz}

\subsubsection{Definition}

While the New Zealand guidance relies on the standard OECD definition for basic AI systems, it specifically introduces and defines Agentic AI in its glossary based on its autonomous capabilities. The framework defines Agentic AI as a system or programme that can ``autonomously perform tasks on behalf of a user or another system by designing its workflow and using available tools.''\footnote{Ministry of Business, Innovation and Employment (MBIE), `Responsible AI Guidance for Businesses' (New Zealand Government, July 2025) 34. Available at: \url{https://www.mbie.govt.nz/assets/responsible-ai-guidance-for-businesses.pdf}} Furthermore, the guidance asserts that the system possesses ``agency'' allowing it to ``make decisions, take actions, solve complex problems, and interact with external environments beyond the data upon which the system's machine learning (ML) models were trained.''\footnote{ibid.}

\subsubsection{Technical Gap}

While the New Zealand framework accurately identifies the mechanical necessity of ``using available tools,''\footnote{ibid.} its definition suffers from profound anthropomorphic friction regarding how these systems actually operate. By asserting that the system uses ``agency'' to ``make decisions'' and ``solve complex problems,''\footnote{ibid.} the guidance attributes conscious, cognitive deliberation to a probabilistic model. More critically, the claim that the system operates by ``designing its workflow''\footnote{ibid.} betrays a structural misunderstanding of the iterative execution loop. From a computer science perspective, an agentic system does not consciously ``design'' a workflow; it generates a probabilistic sequence of text tokens which an external, hardcoded orchestration framework --- such as LangChain or DSPy --- parses into deterministic API calls. By legally defining the AI as a problem-solving entity that ``designs'' its own operational path, the framework invites regulators to govern the illusion of machine intent, rather than properly regulating the programmatic constraints, system prompts, and tool access controls implemented by the human developer.

\subsection{The New South Wales Government}
\label{subsec:nsw}

\subsubsection{Definition}
The NSW Government explicitly separates AI agents from basic automation and standard Generative AI. It defines an AI agent as software that ``identifies tasks to achieve a goal, makes decisions, acts, and adapts continuously through learning.''\footnote{Digital NSW, `AI agent usage and deployment guidance' (NSW Department of Customer Service, October 2025) 13. Available at: \url{https://www.digital.nsw.gov.au/policy/artificial-intelligence/guide-to-using-ai-agents-nsw-government}} Unlike standard AI, the framework asserts that these agents can ``perceive, plan, reason, decide, learn and act autonomously with little or no human input to reach a goal.''\footnote{ibid 4.} Operationally, the framework states that agents ``use models, tools, and data to identify tasks, make decisions, take action and learn over time,''\footnote{ibid 3.} relying on a system Memory defined as ``stored information that helps the agent remember actions, context, and decisions to improve over time.''\footnote{ibid 13.}

\subsubsection{Technical Gap}

The NSW guidance correctly identifies that agents use tools, but it suffers from severe cognitive and architectural friction by attributing human traits to the system's execution loop. The framework's claim that agents ``perceive, plan, reason, decide''\footnote{ibid 4.} falsely elevates probabilistic token prediction to conscious deliberation. More critically, the guidance demonstrates a profound misunderstanding of state management. The claim that an agent ``adapts continuously through learning'' and possesses a ``memory'' to ``remember actions\ldots\ to improve over time''\footnote{ibid 13.} fundamentally misrepresents deployed agentic architectures. Standard agentic deployments use a frozen foundation model whose neural network weights are fixed at deployment and do not update during operation. What the policy misinterprets as ``remembering'' and ``learning'' is simply an iterative execution loop querying a vector database and appending past text strings into its current context window prompt. By legally defining the system as an entity that ``learns'' and ``remembers,'' the framework invites lawmakers to govern the software as an evolving, sentient actor rather than regulating the developer's static orchestration code and database retrieval mechanisms.

\subsection{The Bank of Singapore}
\label{subsec:bos}

\subsubsection{Definition}

The Bank of Singapore describes Agentic AI as representing a ``fundamental shift from passive assistance to proactive autonomy.''\footnote{Bank of Singapore, `Bank of Singapore deploys agentic AI tool to automate writing of source of wealth reports' (Media Release, 10 October 2025). Available at: \url{https://www.bankofsingapore.com/media-releases/2025/bank-of-singapore-deploys-agentic-ai-tool-to-automate-writing-of-source-of-wealth-reports.html}} Unlike standard generative AI, which merely ``reacts to prompts and does not act on its own,'' the Bank asserts that agentic systems ``pursue goals with intelligent initiative.''\footnote{ibid.} The Bank further defines the system's operational mechanics by claiming it can ``independently start tasks, coordinate tools across multiple platforms, adapt dynamically in real time, and continuously enhance their performance through memory and learning,'' progressively refining its behaviour through past interactions.\footnote{ibid.}

\subsubsection{Technical Gap}

The Bank of Singapore's definition illustrates how corporate marketing exacerbates legal friction by anthropomorphising basic software architecture. The claim that its agentic system possesses ``intelligent initiative'' and can ``independently start tasks''\footnote{ibid.} attributes volitional properties to a system that is completely inert until triggered by an explicit programmatic event --- in this case, a relationship manager uploading documents to trigger the underlying script. More critically, the Bank's assertion that the system will ``continuously enhance their performance through memory and learning'' and progressively refine its behaviour\footnote{ibid.} misrepresents the architecture in the same way identified in the NSW guidance above. Deployed foundational models do not learn or alter their neural weights during an execution loop; the model is frozen. What the corporate sector describes as ``learning'' and ``memory'' is Retrieval-Augmented Generation --- a deterministic mechanism that retrieves static text strings from a database and injects them into the current prompt. By framing the AI as a proactive entity that ``learns'' and takes ``initiative,'' corporate communications invite regulators to govern the software as an autonomous, evolving actor rather than focusing on the deterministic API pipelines and database queries executing in the background.

\newpage
\section*{Part~II: Synthesis and Critical Analysis --- The Definitional Gap Between Technical Reality and Legal Construction of Agentic AI}

\section{Introduction to the Synthesis}
\label{sec:intro_synthesis}

The preceding survey of definitions reveals a striking and consistent pattern: across jurisdictions, regulatory cultures, and institutional contexts, no examined legal or policy framework has succeeded in capturing the technical essence of Agentic AI with sufficient precision. The definitions produced by the European Data Protection Supervisor, the European Commission, the EU AI Act, the OECD and G7, NIST, the UK Information Commissioner's Office, the Singapore Personal Data Protection Commission, the New Zealand Ministry of Business Innovation and Employment, the New South Wales Government, and even the private sector --- as exemplified by the Bank of Singapore --- converge, despite their geographic and institutional diversity, upon a remarkably similar set of conceptual errors. These errors are not incidental. They are structural, arising from a fundamental methodological problem: the attempt to regulate a novel, architecturally specific technology using the conceptual vocabulary of either prior AI paradigms or, more troublingly, human cognition and behaviour. This section synthesises those errors into four recurring categories of definitional failure and examines why the non-technical authorship of these frameworks systematically produces them.

\section{The Four Recurring Categories of Definitional Failure}
\label{sec:four_categories}

\subsection{Anthropomorphism: Governing a Mirror of the Human Mind}
\label{subsec:anthropomorphism}

The most pervasive error across the reviewed corpus is what may be termed \textit{definitional anthropomorphism}: the attribution of human cognitive faculties --- reasoning, planning, judgement, decision-making, memory, and learning --- to systems that operate on fundamentally different computational principles.

The European Commission's landscape report defines Agentic AI as possessing ``Reasoning,'' which it elaborates as ``Contextual decision-making to make judgment calls and weigh tradeoffs.''\footnote{European Commission (n~42).} The New South Wales government framework similarly asserts that agents can ``perceive, plan, reason, decide, learn and act autonomously.''\footnote{Digital NSW (n~59).} The UK ICO characterises these systems as capable of ``mimicking some types of reasoning and planning,''\footnote{Information Commissioner's Office (n~31).} while the New Zealand guidance attributes to them an ``agency'' enabling them to ``make decisions'' and ``solve complex problems.''\footnote{Ministry of Business, Innovation and Employment (MBIE) (n~54).} The Singapore Model AI Governance Framework, at its very foundation, defines AI as technology that seeks to ``simulate human traits such as knowledge, reasoning, problem solving, perception, learning and planning.''\footnote{PDPC (n~47).}

This convergence on cognitive language is not trivial. It produces a regulatory object that does not exist: an autonomous, deliberating agent capable of spontaneously forming intentions. The EDPS exemplifies this most acutely in its warning that Agentic AI might ``autonomously determine new uses for personal data''\footnote{Andy Goldstein (n~1).} --- a formulation that treats the AI as a sentient employee capable of reconsidering its own mandate. As the G7 Code of Conduct similarly warned against systems being ``self-replicating or able to train other models,''\footnote{White \& Case (n~20).} the underlying metaphor is biological and cognitive, not computational. From an engineering standpoint, a Large Language Model does not deliberate; it executes probabilistic token prediction across a high-dimensional vector space to minimise a loss function. It does not ``weigh tradeoffs'' in any meaningful cognitive sense --- it generates the statistically most probable next token given a system prompt and an input. An agent's behaviour is strictly bounded by the system prompt, the tool permissions granted to it by the developer, and the programmatic constraints of its orchestration framework. It cannot, by definition, spontaneously ``determine new uses'' for data that fall outside the scope of its programmed objective function unless the developer architecturally failed to implement adequate access controls.

This matters for law not merely as an abstract technical quibble. If regulators define Agentic AI as a system that ``reasons'' and ``decides,'' they direct their compliance regimes toward governing the system's internal outputs --- its ``judgement calls'' --- rather than the deterministic code, Identity and Access Management permissions, and API boundaries that actually govern its external actions. The regulatory target becomes an anthropomorphic fiction rather than a set of auditable engineering decisions.

\subsection{Architectural Conflation: Mistaking the Number of Agents for the Nature of Agency}
\label{subsec:conflation}

A second major category of error, appearing in both the EDPS guidance and the European Commission's landscape report, concerns the conflation of Agentic AI with Multi-Agent Systems (MAS) --- the mistaken belief that agentic behaviour is defined by a plurality of communicating sub-agents rather than by a specific execution architecture.

The EDPS defines Agentic AI specifically as an orchestrator that ``coordinating multiple agents, managing their communication, and distributing tasks to accomplish larger, more complex objectives,''\footnote{Andy Goldstein (n~1).} while the European Commission asserts definitively that Agentic AI is ``a multi-agent system in which each agent performs a specific subtask required to reach a goal, with their efforts coordinated through AI orchestration.''\footnote{European Commission, `European agentic AI landscape' (n~42) 4.} Both definitions imply that a single AI system, however autonomously it operates, cannot be ``agentic'' unless it is directing multiple subordinate agents.

This is a categorical architectural error. In computer science, ``agentic'' is a behavioural property defined by the presence of a continuous perception-reasoning-action loop --- such as the ReAct (Reason, Act, Observe) framework --- combined with the capacity to invoke external tools and APIs. A single LLM operating in such a loop, granted write-access to a live database and the ability to execute code, is highly agentic, regardless of whether it communicates with any other models. Conversely, a sophisticated Multi-Agent System in which multiple models pass messages to one another without any external tool-calling or iterative execution loop may exhibit very limited real-world agency. By defining the category by the headcount of communicating models rather than by the architectural mechanism of the execution loop and tool integration, these frameworks draw a regulatory boundary around the wrong feature of the technology. An AI system's capacity to cause real-world harm --- to access personal data, execute financial transactions, modify files, or interact with external infrastructure --- is a function of its tool permissions and iterative loop architecture, not of how many language models happen to be involved.

\subsection{The ``Levels of Autonomy'' Fallacy: Treating a Structural Shift as a Degree}
\label{subsec:autonomy_fallacy}

A third recurring error, prominent in the EU AI Act, NIST, and the Singapore framework, is the conceptualisation of autonomy as a scalar property --- a degree or ``level'' on a continuous spectrum --- rather than as an architectural discontinuity.

The EU AI Act defines an AI system as one designed to operate with ``varying levels of autonomy,''\footnote{Regulation (EU) 2024/1689 (n~14).} while NIST describes AI systems as ``designed to operate with varying levels of autonomy,'' noting that ``human-AI configurations can span from fully autonomous to fully manual.''\footnote{NIST (n~23).} The OECD framework, on which the G7 Guiding Principles directly build, similarly assumes these systems ``vary in their levels of autonomy and adaptiveness after deployment.''\footnote{OECD (n~18).} Singapore's framework categorises agentic behaviour under a ``Human-out-of-the-loop'' designation, defining this as a situation where ``there is no human oversight over the execution of decisions'' and the ``AI system has full control without the option of human override.''\footnote{Personal Data Protection Commission Singapore (PDPC) (n~47) 30.}

The practical legal consequence of this framing is that it places a heavily prompted conversational LLM and a fully autonomous API-calling agentic system on the same regulatory spectrum, differentiated only by the degree to which a human is present. This misses the fundamental engineering distinction between the two. The shift from a standard LLM to an agentic system is not achieved by reducing human presence; it is achieved by wrapping the model in an iterative execution loop --- a programmatic \texttt{while} loop that triggers the model to generate text, parses that text as a function call, executes the function against an external API, feeds the result back into the model's context window, and repeats the cycle until a termination condition is met. This is a structural, architectural difference in kind, not a difference in degree of human absence.

Furthermore, Singapore's formulation that the ``AI system has full control''\footnote{Ibid.} in a human-out-of-the-loop scenario incorrectly anthropomorphises the concept of control. No AI system possesses ``control'' in a deliberative sense. Autonomy is not freedom of will; it is the absence of a human confirmation step between the model's text output and the execution of a function call by the surrounding programmatic infrastructure. Regulators who conceptualise this as the AI having ``full control'' direct their compliance frameworks toward governing the model's hypothetical agency rather than auditing the developer's explicit decisions about which function calls to permit, which APIs to expose, and which sandboxing constraints to enforce around the execution loop.

\subsection{The ``Learning and Memory'' Myth: Mischaracterising State Management as Continuous Adaptation}
\label{subsec:learning_myth}

A fourth category of error, present in the NSW Government framework, the Bank of Singapore's corporate communication, and the New Zealand guidance, concerns a systematic misrepresentation of state management mechanisms --- specifically, the conflation of Retrieval-Augmented Generation and context-window injection with genuine, continuous machine learning.

The NSW Government framework claims that agents ``adapt continuously through learning'' and possess a ``memory'' to ``remember actions, context, and decisions to improve over time.''\footnote{Digital NSW (n~59) 13.} The Bank of Singapore similarly describes its agentic system as able to ``continuously enhance their performance through memory and learning,'' progressively refining its behaviour through past interactions.\footnote{Bank of Singapore (n~65).} The New Zealand guidance holds that an agent operates ``beyond the data upon which the system's machine learning (ML) models were trained,''\footnote{Ministry of Business, Innovation and Employment (MBIE) (n~54) 34.} implying an ongoing, dynamic expansion of the model's knowledge base.

These characterisations fundamentally misrepresent the architecture of deployed agentic systems. Standard agentic deployments use a frozen foundation model --- that is, a model whose neural network weights are fixed at the time of deployment and do not update during operation. What these frameworks describe as ``learning'' and ``memory'' is, technically, one of two deterministic operations: either the injection of prior interaction logs as text strings into the model's context window at the start of each new execution cycle, or the retrieval of stored text documents from a vector database via Retrieval-Augmented Generation and their inclusion in the current prompt. Neither operation constitutes learning in any machine learning sense; the model's parameters remain unchanged. The system's outputs may appear to reflect prior interactions, but this is because relevant text from those interactions is being fed back into the input, not because the model has updated its weights.

The legal consequences of this mischaracterisation are significant. If a framework defines an agentic system as one that ``learns'' and ``adapts'' over time, it may prompt regulatory requirements --- such as re-certification, impact assessments, or re-registration --- triggered by the AI's perceived autonomous evolution, when in reality the system has not changed at all. The regulatory burden would be calibrated to govern a dynamic, self-modifying actor that does not exist, while potentially failing to scrutinise the static but consequential developer decisions about what data to store, what the retrieval mechanism surfaces, and what the system prompt instructs the model to do with retrieved content.

\section{The Structural Reasons for Definitional Failure: Why Non-Technical Authors Produce These Errors}
\label{sec:structural_reasons}

The four categories of error identified above are not random. They reflect identifiable structural conditions in the production of legal and policy definitions of emerging technologies.

\subsection{The Absence of Interdisciplinary Authorship}
\label{subsec:absence_interdisciplinary}

The most fundamental reason for the definitional failures surveyed here is that the frameworks were produced without meaningful participation from computer scientists or AI engineers at the drafting stage. Legal and policy instruments are authored by lawyers, civil servants, diplomats, and policy analysts whose professional formation equips them to regulate human behaviour, institutional relationships, and social risks, but not to characterise the internal mechanics of probabilistic computational systems. No amount of policy expertise compensates for the absence of the technical understanding required to distinguish between a model's text output and the deterministic code that acts upon it, or between a context window injection and a genuine update to neural network weights.

This structural absence of technical authorship is not a criticism of individual competence. It reflects an institutional design problem: the processes by which these frameworks were produced --- parliamentary committees, inter-agency working groups, public consultations --- do not routinely integrate deep technical expertise at the drafting stage. The OECD definition of an AI system as a system that ``infers, from the input it receives, how to generate outputs''\footnote{OECD (n~18).} is an adequate description of a standard predictive model, but it was drafted before the architectural shift to agentic systems had matured, and it has been carried forward by the G7 and other frameworks without the technical revision its limitations now require.

\subsection{The Seductive Availability of Human Cognitive Vocabulary}
\label{subsec:seductive_vocabulary}

A second structural reason for anthropomorphism across these frameworks is the absence of an alternative vocabulary. When policymakers seek to describe what an AI system does, the only readily available conceptual toolkit is one developed to describe human and animal behaviour: reasoning, deciding, planning, learning, remembering. These terms are deeply embedded in legal systems --- human decision-makers are routinely evaluated on the quality of their ``reasoning'' and ``judgment'' in administrative and tort law --- and they transfer naturally to descriptions of AI, particularly when AI developers themselves use this vocabulary in product communications.

The Bank of Singapore's description of its system as pursuing ``goals with intelligent initiative''\footnote{Bank of Singapore (n~65) 16.} is a corporate marketing document, but it is exactly the kind of language that non-technical policymakers encounter and absorb. When the commercial sector, the academic literature, and popular discourse all describe AI systems as ``reasoning'' and ``deciding,'' it becomes epistemically costly for a regulator to insist on more technically precise --- but less intuitive --- formulations. The result is a regulatory vocabulary borrowed wholesale from human cognition and applied, uncritically, to a probabilistic text-generation architecture.

\subsection{The Inherited Inadequacy of Prior AI Regulatory Concepts}
\label{subsec:inherited_inadequacy}

A third structural reason lies in the path-dependency of regulatory frameworks. Many of the definitions reviewed here --- particularly the OECD's, NIST's, and the EU AI Act's --- were developed during an earlier phase of AI development, when AI systems were primarily understood as predictive or classifying tools: systems that took an input and produced an output. The concept of ``varying levels of autonomy'' made sense in this context: one could imagine a spectrum from a fully automated classification system with no human review to a system whose outputs were always reviewed before action was taken.

The emergence of agentic architectures --- systems that do not produce a single output but instead engage in an iterative, multi-step execution loop that interfaces with live external environments --- has rendered these inherited concepts structurally inadequate. Yet no framework has yet conducted the foundational review required to replace them. The EU AI Act's ``intended purpose'' framework,\footnote{Regulation (EU) 2024/1689 (n~14) 12.} for example, was designed around a static, discernible deployment objective against which compliance could be assessed. When applied to an agentic system whose execution paths are dynamically generated and non-deterministic, the concept breaks down: the ``purpose'' of the system is known (achieve a goal), but the specific actions it will take to achieve that purpose cannot be predicted, certified, or legally pre-specified in the way the Act envisages. Similarly, the Act's rigid division of liability between a ``provider'' --- the creator of the foundational model --- and a ``deployer'' --- the entity using it\footnote{ibid.} --- cannot accommodate the composite architecture of an agentic system, in which the ``agency'' of the system emerges not from the foundational model alone but from the combination of the model, the orchestration framework wrapping it, and the tool permissions granted to it. A deployer who takes an open-weight model, wraps it in a LangChain execution loop, and grants it write access to a customer database has done something architecturally transformative to the system's agency, yet the Act provides no clear framework for attributing liability to this composite construction.

\subsection{The Absence of Binding Legal Definitions Creates a Definitional Vacuum Filled by Policy Approximations}
\label{subsec:definitional_vacuum}

A final structural observation is that, as the EDPS itself implicitly acknowledges by positioning its definition as a technical and policy description rather than a binding legal instrument,\footnote{Andy Goldstein (n~1).} and as the UK ICO explicitly recognises in conceding that there is ``no legal definition of agentic AI'',\footnote{Slaughter and May (n~28).} the definitional field is currently occupied not by authoritative legal norms but by a proliferation of non-binding approximations. These approximations --- produced by data protection authorities, risk management bodies, government guidance documents, and corporate communications --- have varying degrees of technical accuracy and no binding normative force. The result is a definitional landscape characterised by significant inconsistency: the EDPS requires multiple agents; the New Zealand framework requires only one.\footnote{Ministry of Business, Innovation and Employment (MBIE) (n~54) 34.} There is overlapping terminological confusion --- the EU AI Act would classify an agentic system simply as an ``AI system'' subject to its risk-based tiers,\footnote{Regulation (EU) 2024/1689 (n~14).} while NIST would treat it as a standard automated system operating at a certain ``level'' of autonomy\footnote{NIST (n~23) 1.} --- and a systematic tendency toward the errors analysed above.

\section{Conclusion: The Implications of the Definitional Gap}
\label{sec:conclusion}

Taken together, the definitional failures surveyed in this section reveal not a failure of regulatory will but a failure of regulatory epistemology. The frameworks examined here were not produced by negligent or indifferent actors; they were produced by competent legal and policy professionals working within institutional structures that were not designed to integrate the kind of technical expertise that accurate definition of Agentic AI requires. The result is a corpus of definitions that, while varying in their specific errors, converge upon a common analytical misstep: they attempt to regulate the appearance of agentic behaviour --- the cognitive-sounding outputs of a probabilistic text-generation system --- rather than the mechanisms that produce and constrain that behaviour: the iterative execution loop, the tool permissions, the system prompt, the sandboxing architecture, and the access controls implemented by the developer.

The consequences of this misalignment are not merely academic. When legal frameworks regulate the AI's apparent ``reasoning'' rather than the developer's architectural decisions, they generate compliance obligations that are simultaneously overbroad --- targeting any AI system that uses cognitive-sounding language to describe its outputs --- and underinclusive, failing to reach the specific engineering decisions about tool access, data permissions, and execution loop design that actually determine the real-world risk profile of an agentic deployment. The path toward more adequate regulation does not require the abandonment of existing frameworks; it requires their revision in close collaboration with the engineering community that designs and builds the systems those frameworks seek to govern.

\newpage
\section*{Bibliography}
\label{sec:bibliography}

\begin{list}{}{
    \setlength{\leftmargin}{2em}
    \setlength{\itemindent}{-2em}
    \setlength{\itemsep}{0.5em}
}

\item Regulation (EU) 2024/1689 of the European Parliament and of the Council of 13 June 2024 laying down harmonised rules on artificial intelligence [2024] OJ L 2024/1689 (EU AI Act)

\item Anthropic, 'Empirical Work on Agents' (Anthropic Research, 2026) <\url{https://www.anthropic.com/research/measuring-agent-autonomy}>

\item Bank of Singapore, 'Bank of Singapore deploys agentic AI tool to automate writing of source of wealth reports' (Media Release, 10 October 2025) <\url{https://www.bankofsingapore.com/media-releases/2025/}>

\item C3.ai, 'Agentic System Architecture' (C3.ai Documentation, 2026) <\url{https://c3.ai/blog/agentic-ai-explained/}>

\item Darling C, Tobey D and Carr A, 'G7 publishes guiding principles and code of conduct for artificial intelligence' (DLA Piper, 2023)

\item Digital NSW, \textit{AI agent usage and deployment guidance} (NSW Department of Customer Service, October 2025) <\url{https://www.digital.nsw.gov.au/policy/artificial-intelligence/guide-to-using-ai-agents-nsw-government}>

\item European Commission, \textit{European agentic AI landscape} (2025) <\url{https://digital-strategy.ec.europa.eu/en/library/agentic-ai-leveraging-european-ai-talent-and-regulatory-assets-scale-adoption}>

\item Goldstein A, 'Agentic AI' (European Data Protection Supervisor, TechSonar, 2026) <\url{https://www.edps.europa.eu/data-protection/technology-monitoring/techsonar/agentic-ai_en}>

\item Google Cloud, 'Agentic AI' (Google Cloud Documentation, 2026) <\url{https://cloud.google.com/discover/what-is-agentic-ai}>

\item Information Commissioner's Office, 'AI'll get that! Agentic commerce could signal the dawn of personal shopping AI-gents' (ICO News and Blogs, 8 January 2026) <\url{https://ico.org.uk/about-the-ico/media-centre/news-and-blogs/2026/01/ai-ll-get-that/}>

\item Maas MM and Olasunkanmi T, 'Treaty-Following AI' (2025) Institute for Law \& AI Working Paper Series No. 1-2025 <\url{https://law-ai.org/wp-content/uploads/2025/12/Maas-Olasunkanmi-2025-Treaty-Following-AI-final-LawAI-WPS-1-2025-4.pdf}>

\item Meta, 'AI Security Research' (Meta AI Research, 2026) <\url{https://ai.meta.com/blog/practical-ai-agent-security/}>

\item Microsoft, 'Agent Factory: The new era of agentic AI' (Microsoft Azure, 2026) <\url{https://azure.microsoft.com/en-us/blog/agent-factory-the-new-era-of-agentic-ai-common-use-cases-and-design-patterns/}>

\item Ministry of Business, Innovation and Employment, \textit{Responsible AI Guidance for Businesses} (New Zealand Government, July 2025) <\url{https://www.mbie.govt.nz/assets/responsible-ai-guidance-for-businesses.pdf}>

\item National Institute of Standards and Technology, \textit{Artificial Intelligence Risk Management Framework (AI RMF 1.0)} (NIST AI 100-1, US Department of Commerce, January 2023) <\url{https://nvlpubs.nist.gov/nistpubs/ai/NIST.AI.100-1.pdf}>

\item OECD, 'Recommendation of the Council on Artificial Intelligence' (OECD/LEGAL/0449, 2024) <\url{https://legalinstruments.oecd.org/en/instruments/OECD-LEGAL-0449}>

\item OpenAI, 'Developer Documentation on Agents' (OpenAI Developer Platform, 2026) <\url{https://developers.openai.com/tracks/building-agents/}>

\item Personal Data Protection Commission Singapore, \textit{Model Artificial Intelligence Governance Framework} (2nd edn, 2020) <\url{https://www.pdpc.gov.sg/-/media/files/pdpc/pdf-files/resource-for-organisation/ai/sgmodelaigovframework2.pdf}>

\item Slaughter and May, 'ICO publishes report on agentic AI and its data privacy implications' (The Lens, 2026) <\url{https://thelens.slaughterandmay.com/post/102me9n/}>

\item White \& Case, 'AI Watch: Global regulatory tracker --- G7' (4 December 2024) <\url{https://www.whitecase.com/insight-our-thinking/ai-watch-global-regulatory-tracker-g7}>

\end{list}

\end{document}